# Enhanced spontaneous emission in a photonic crystal light-emitting diode

M. Francardi[(1,4)]*, L. Balet[(2,3)]*, A. Gerardino[(1)], N. Chauvin[(2)], D. Bitauld[(2)], L.H. Li[(3)], B. Alloing[(3)], and A. Fiore[(2)]

(1) Institute for Photonics and Nanotechnologies-CNR, via Cineto Romano 42, 00156 Roma, (Italy)

(2) COBRA Research Institute, Eindhoven University of Technology, PO Box 513, 5600MB Eindhoven, The Netherlands

(3) Ecole Polytechnique Fédérale de Lausanne (EPFL), Institute of Photonics and Quantum Electronics, CH-1015 Lausanne (Switzerland)

(4) Email: marco.francardi@ifn.cnr.it

* These authors contributed equally to this work.

ABSTRACT. We report direct evidence of enhanced spontaneous emission in a photonic crystal (PhC) light-emitting diode. The device consists of p-i-n heterojunction embedded in a suspended membrane, comprising a layer of self-assembled quantum dots. Current is injected laterally from the periphery to the center of the PhC. A well-isolated emission peak at 1.3 µm from the PhC cavity mode is observed, and the enhancement of the spontaneous emission rate is clearly evidenced by time-resolved electroluminescence measurements, showing that our diode switches off in a time shorter than the bulk radiative and nonradiative lifetimes.

Corresponding author: Marco Francardi, phone: +390641522242, e-mail : marco.francardi@ifn.cnr.it

Spontaneous emission results from the coupling of a quantum emitter to the surrounding electromagnetic environment. It therefore depends both on the emitter's dipole moment, and on the density and properties of the available electromagnetic modes[1]. In particular, a wavelength-scale, low-loss cavity can provide an increase of spontaneous emission rate by a factor $\propto \frac{Q}{V}$ (Q: quality factor, V: mode volume) over the free-space value. This effect can be used to change the emission rate and at the same time efficiently funnel most of the spontaneously-emitted photons into a single, well-controlled output mode, resulting in high extraction efficiency. After the first demonstration of enhanced spontaneous emission from quantum dots (QDs) in monolithic micropillars[4], several important steps have been realized, such as the demonstration of inhibited spontaneous emission[5] and of coupling of single QDs to the cavity mode[6]. However, all these experiments have used optical pumping to excite the emitter. For practical applications, electrical pumping is needed. The implementation of electrical injection in a low-loss, ultrasmall cavity poses tremendous fabrication and experimental challenges: On one hand, electrical contacts must be integrated in a micrometer-scale device without significantly increasing the optical loss. On the other hand, the experimental demonstration and practical application of enhanced spontaneous emission requires an ultrafast (sub-ns) electrical probing, requiring a careful control of device parasitics. Indirect evidence of spontaneous emission control in electroluminescence has been recently deduced from the static characteristics of metallic-coated nanolasers[9] and of electrically-contacted micropillars[10]. In this paper we report a high-frequency PhC LED structure which allows the direct experimental measurement of cavity-enhanced spontaneous emission dynamics in a light-emitting diode. Furthermore, the emission wavelength is around 1.3 μm, corresponding to a transmission window of optical fibers.

Our device (fig. 1) consists of a membrane photonic crystal cavity containing a p-i-n heterojunction, where holes are injected from a top ring p-contact, and electrons from the sides of the mesa, using highly-doped GaAs contact layers to spread the current throughout the mesa to the center of the cavity. As compared to injection through a central post[12], this approach allows an easier fabrication, and the

Corresponding author: Marco Francardi, phone: +390641522242, e-mail : marco.francardi@ifn.cnr.it

decoupling of the problems of electrical injection and optimisation of quality factor. Additionally, in contrast to previous approaches to PhC membrane LEDs[11,12], we are able to electrically address a single PhC cavity with low parasitic resistance and capacitance, allowing the fast electrical control needed for the time-resolved measurement of the emission dynamics.

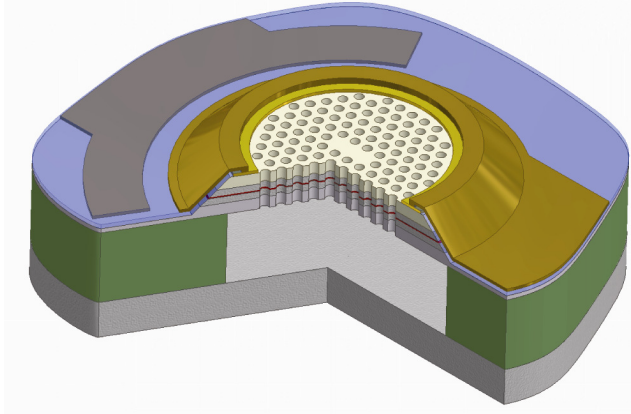

**Figure 1**. Sketch of the LED showing the p-i-n heterojunction suspended membrane.

The heterostructure used to fabricate the LEDs is grown by molecular beam epitaxy (MBE) on a GaAs substrate and consists of a 370 nm-thick GaAs/Al$_{0.2}$Ga$_{0.8}$As p-i-n junction on top of a 1500 nm-thick Al$_{0.7}$Ga$_{0.3}$As sacrificial layer. A single layer of low-density InAs QDs (5-7dots /μm$^2$) emitting at 1300 nm at 5K is grown using a very low InAs deposition rate ≈2x10$^{-3}$ monolayers/s.[13] The fabrication process is based on several e-beam lithography steps and typical thin-film processes. We first define 8 μm-diameter, 320 nm-deep circular mesas by wet etching down to the bottom n-contact layer. Then, a ring-shaped n-contact is defined by lift-off of a 155nm-thick Ni/Ge/Au/Ni/Au multilayer that is subsequently annealed at 400°C for 30 min. A 200 nm-thick Si$_3$N$_4$ layer is deposited by plasma-enhanced chemical vapor deposition (PECVD) to isolate the n and p contacts. The Si$_3$N$_4$ is then removed from the n-contact and from the top of the mesa by selective reactive ion etching (RIE). By lift-off of a Cr/Au (thickness: 110 nm) bilayer we realize an annular p-contact on the top of the mesa. In the same evaporation step, the top p-contact and the bottom n-contact are connected to ground-signal-ground coplanar electrodes on top of the Si$_3$N$_4$ surface, for contacting with a high frequency probe.

Corresponding author: Marco Francardi, phone: +390641522242, e-mail : marco.francardi@ifn.cnr.it

Three tilted Cr/Au evaporations are performed in order to obtain a continuous film over the mesa lateral edge. Finally, an annular gold cover is evaporated around the entire mesa edge in order to filter out light scattered by the mesa sidewalls. To integrate the PhC nanocavity on the LED, a 150 nm-thick $SiO_2$ layer is then deposited and the PhC pattern – aligned to the mesa - is defined on a 200 nm-thick resist and transferred to the $SiO_2$ layer and then to the GaAs/AlGaAs membrane ($SiCl_4/O_2/Ar$ based RIE). Finally, the bottom $Al_{0.7}Ga_{0.3}As$ sacrificial layer is selectively removed using a diluted HF solution. A scanning-electron microscope (SEM) image of the LED at the end of the fabrication process is shown in Fig 2. The PhC nanocavities that have been integrated with the LED are L3 modified nanocavities[1,5], characterized by a triangular lattice with a fixed filling factor (f = 35%) and a variable lattice parameter a ( from 292nm to 352nm with a step of 10 nm).

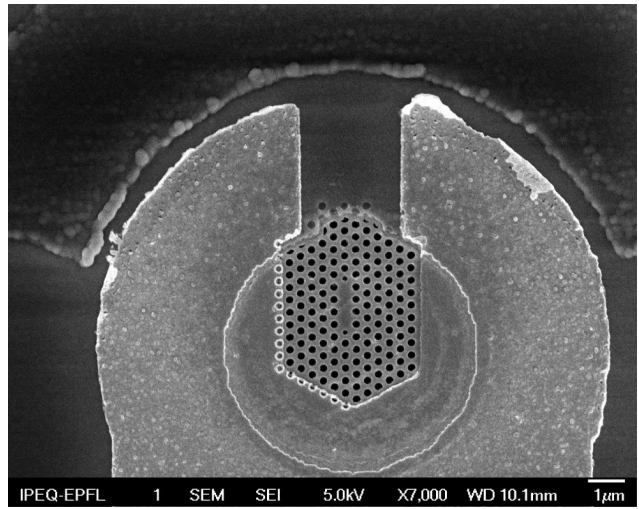

**Figure 2.** Scanning Electron Micrograph of a 8 µm-diameter PhC LED. The p-contact is clearly visible on the bottom side, whereas the n contact is shaded by the $Si_3N_4$ insulating layer.

The device testing was performed using a cryogenic probe station (T=5K) equipped with cooled 50Ω microwave probes. The LED current-voltage characteristic is shown in Fig. 3. A clear diode characteristics is observed even at low temperature, with forward-bias (reverse-bias) threshold voltage around 2V (-7V), indicating very good electrical contacts.

Corresponding author: Marco Francardi, phone: +390641522242, e-mail : marco.francardi@ifn.cnr.it

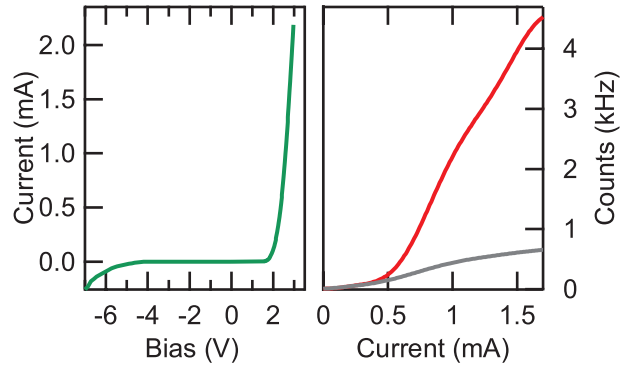

**Figure 3.** (left) IV characteristic of the device exhibiting diode operation. (right) Electroluminescence as a function of the injected current at the wavelength of the mode (red curve) and detuned from the mode (grey curve).

Light emitted from the device was collected from the top using an external Cassegrain microscope objective (numerical aperture NA=0.4), and refocussed to a single-mode fiber, providing a 1.8-μm diameter collection spot size. The electroluminescence (EL) spectrum at I=1.25 mA is shown in Fig. 4 (red line).

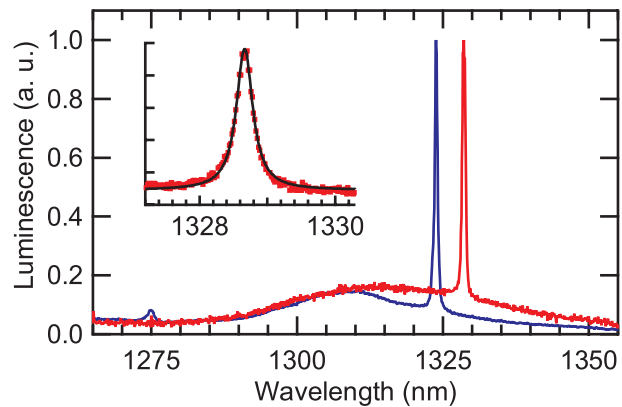

**Figure 4.** Photoluminescence (blue, 750 nm, 80 MHz) and electroluminescence (red, 2.3 V dc, 40 MHz) spectra of a PhC L3 cavity with lattice constant a=341 nm (mesa diameter: 10 μm). The quality factor of the mode is Q≈4600 (inset).

A clear cavity mode is observed around 1329 nm, superposed on a broad emission line corresponding to ground-state emission from the QD ensemble (due to the limited spatial resolution of the set-up,

Corresponding author: Marco Francardi, phone: +390641522242, e-mail : marco.francardi@ifn.cnr.it

single QD lines are not resolved in electroluminescence). The full-width half-maximum (FWHM) of the mode peak is 0.29 nm (see inset in Fig. 4) corresponding to a quality factor Q= 4600. The emission spectrum under optical pumping, measured on the same, open-circuited device in a different microphotoluminescence set-up, presents a very similar Q factor of ≈4000 (blue line). The small red-shift (≈5 nm) observed in the EL spectrum is likely due to different effect of adsorption of residual gas molecules in the two cryostats, and/or to higher device temperature under electrical pumping. The total emission intensity in the mode was filtered with a tunable bandpass filter with a FWHM of 0.2 nm and measured with a fiber-coupled superconducting single photon detector (SSPD)[14]. The mode intensity is plotted as a function of injection current in Fig. 3 (right part, red line), along with the intensity measured when the filter is detuned by 15 nm from the mode (grey line). At low currents <0.5 mA, the two curves are superposed, and indeed no emission peak from the mode is observed in the spectra (not shown). The mode peak appears at currents larger than 0.5 mA, and becomes much more intense than the QD ensemble. Despite the threshold-like behavior of the light-current characteristics, we can rule out the occurrence of lasing since the threshold modal gain is much larger than the gain expected from the low-density QD ensemble, and the apparent threshold almost disappears when the temperature is increased to 77 K. Instead, the nonlinear light-current characteristics are due to the nonuniform current injection in the mesa: At low current carriers are injected only around the mesa edge (as confirmed by spatial imaging the EL), this effect being more evident at low temperatures due to lower hole conductivity.

In order to provide a direct evidence of the spontaneous emission enhancement, we performed time-resolved (TR) EL measurements on a cavity mode (on-resonance) at 1319 nm with a quality factor Q=1000 and off-resonance at 1310 nm. A constant bias voltage with a value $V_B$= 2.3 V, below the threshold of the LED, is applied to the cavity. A TTL signal is added on this constant voltage with a 40 MHz frequency and a 0.6 V amplitude, which results in the waveform shown in the inset of Fig. 5 and produces a periodic switching of the diode. The falling edge of this signal is used to measure the decay time of emission from the LED. The measured average current is 1 mA. The EL signal is spectrally filtered by bandpass filters with a FWHM of 0.2 nm (cavity mode) and 0.8 nm (off resonance emission),

Corresponding author: Marco Francardi, phone: +390641522242, e-mail : marco.francardi@ifn.cnr.it

and coupled to the SSPD. The SSPD output pulse and the TTL trigger are sent to a correlation card to produce a histogram of photon arrival times. The set-up temporal resolution is determined by measuring the temporal decay (Fig. 5, blue diamonds) of emission from the wetting layer, whose lifetime is expected to be well below 100 ps. The measured decay time of ≈210 ps is mainly determined by the jitter of the SSPD and of the correlation card, as previously measured[16], which confirms that device parasitics are low and do not limit the dynamic behavior. The temporal decays of the cavity mode (red squares) and QD ensemble (blue dots) are shown in Fig. 5.

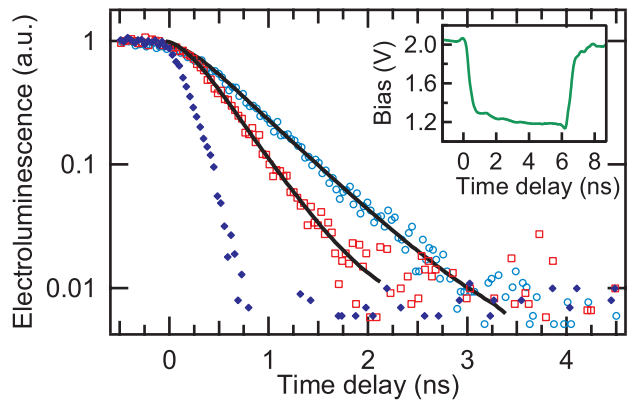

**Figure 5.** Time resolved electroluminescence of the wetting layer (diamonds), mode (open squares) and detuned from the mode (open circles), under 40 MHz pulsed excitation. The electrical pulse waveform is shown in the inset.

The time constants of both curves are then calculated by fitting the experimental results using the convolution of a monoexponential curve with the time response of the experimental setup (fits are shown as lines in Fig. 5). Recombination lifetimes of 380 ps and 580 ps are obtained for the on-resonance and off-resonance emissions respectively. Thus, an increase of the recombination rate by a factor 1.5 is observed confirming a Purcell effect. We note that the off-resonance emission is faster than the typical exciton lifetime of similar QDs in bulk which is around 1 ns[16]. This broad emission peak is attributed to the combination of multiexcitonic emission lines from several QDs, which was already observed to decay faster than single excitons[17]. The theoretical value of the Purcell effect in this cavity,

Corresponding author: Marco Francardi, phone: +390641522242, e-mail : marco.francardi@ifn.cnr.it

for an ideal emitter spectrally narrower than the cavity and spatially aligned to the cavity mode, is $F_p = 3Q\lambda^3 / 4\pi^2 n^3 V_{eff} \approx 100$. Due to large linewidth of the broad peak ($\Delta\lambda \approx 10$ nm), the emission rate enhancement $f_E$ is modified to $f_E = F_p /(1 + 2Q\gamma) \approx 6$ with $\gamma = \Delta\lambda / \lambda$ where $\lambda$ is the wavelength of the emitter[20]. The lower experimental value is attributed to the unoptimised spatial coupling of the QDs to the cavity mode. By reducing the injection area it should be possible to isolate the EL of single QDs and achieve higher Purcell factors.

In conclusion, we have demonstrated the enhancement of spontaneous emission dynamics in a photonic-crystal quantum dot LED. On one hand, this opens the way to the fabrication of LEDs, diode lasers and single-photon sources with a modulation speed and turn-on delay not limited by the free-space spontaneous emission time. On the other hand, the integration of electrical contacts with a coupled QD-cavity system allows the ultrafast electrical control of the coupling (for example through the Stark tuning of the exciton energy), on a time scale shorter than the exciton coherence time. This will allow the coherent manipulation of excitonic and photonic qubits, a major step forward in the field of solid-state cavity quantum electrodynamics.

ACKNOWLEDGMENT. We are thankful to F. Römer and B. Witzigmann (ETHZ) for fruitful discussions, and we acknowledge funding from the Swiss State Secretariate for Education and Research through COST-P11 Project No. C05-70, the Swiss National Science Foundation through the "Professeur boursier" and NCCR Quantum Photonics program, EU-FP6 IP "QAP" Contract No. 15848, the Network of Excellence "ePIXnet," and the Italian MIUR-FIRB program.

Corresponding author: Marco Francardi, phone: +390641522242, e-mail : marco.francardi@ifn.cnr.it

Corresponding author: Marco Francardi, phone: +390641522242, e-mail : marco.francardi@ifn.cnr.it

Corresponding author: Marco Francardi, phone: +390641522242, e-mail : marco.francardi@ifn.cnr.it